\documentclass[12pt,amssymb]{article} 

\newcommand{\be}{\begin{equation}}
\newcommand{\ee}{\end{equation}}
\newcommand{\bea}{\begin{eqnarray}}
\newcommand{\eea}{\end{eqnarray}}

\begin{document}

\begin{titlepage}
\begin{center}
\vskip .2in \hfill \vbox{
    \halign{#\hfil         \cr
           hep-th/0204119 \cr
           UCSD-PTH-02-05\cr
           April 2002    \cr
           }  
      }   
\vskip 1.5cm {\Large \bf A Brief Comment on Instanton-like
               Singularities and Cosmological Horizons} \\
\vskip .1in
\vskip .3in {\bf  Jason Kumar}
\footnote{e-mail address:j1kumar@ucsd.edu}\\
\vskip .25in {\em  Department of Physics,
University of California, San Diego\\
La Jolla, CA  92093-0354 USA \\} \vskip .1in \vskip 1cm
\end{center}
\begin{abstract}
We argue that in the presence of instanton-like singularities, the
existence of cosmological horizons can become frame-dependent,
ie. a horizon which appears in Einstein frame may not appear
in string frame.  We speculate on the relation
between instanton-like singularities and the formulation of quantum
gravity in de Sitter space.
\vskip 0.5cm
\end{abstract}
\end{titlepage}
\newpage

\section{Introduction}

Recently, one of the most interesting questions to be
examined by string theory is the question of how one formulates
quantum gravity in de Sitter space
(see \cite{Spradlin:2001pw}, and references therein).
In quantum gravity,
it is not yet clear how one can define any gauge-invariant
quantities aside from the S-matrix.  However, de Sitter
space has cosmological horizons which seem to make it
impossible to define an S-matrix.  This problem seems to
be made more interesting by recent astronomical observations
which are consistent with a positive cosmological constant.
And it has been shown (\cite{Hellerman:2001yi}
\cite{Fischler:2001yj}) that even if the cosmological
constant is in fact zero, a universe with eternally accelerating
expansion (as in some quintessence models)
will still exhibit cosmological horizons which make the
formulation of quantum gravity problematic.

It is important to first realize that this is not truly
a phenomenological problem.
Certainly there is a sense in which the scattering amplitudes
computed to explain everything from accelerator experiments to
planetary motion are reasonable approximations, despite the
potential presence of cosmological horizons.  In any case,
current astronomical experiments
are perfectly consistent with a universe whose expansion will
eventually stop accelerating at some time in the future.  In
such a universe there are no cosmological horizons, and thus no
obstruction to the defining quantum gravity in terms of an
S-matrix as a formal matter of principle.

Secondly, it is important to note that the question of cosmological
horizons does not arise only in string theory.  It appears to be
a general statement about quantum gravity that the only well-defined
gauge-invariant (and hence diffeomorphism-invariant) quantities that
are well-understood are S-matrices, which seem problematic in the
presence of cosmological horizons.

But despite the fact that these cosmological horizons do not cause
a phenomenological problem, they do raise an interesting theoretical
and aesthetic question, namely, does the mere existence of quantum
gravity require that the expansion of the universe must stop
accelerating at some point in the future?

A very important development related to this problem is the recent
study of instanton-like singularities
\footnote{We are grateful to M. Costa for pointing out an unfortunate
error in notation in an earlier version of this work.}
\cite{Khoury:2001bz}
\cite{Balasubramanian:2002ry} \cite{Gutperle:2002ai}
(see also \cite{Cornalba:2002fi}, for a related discussion).
We will argue that in the
presence of such singularities, the transition between Einstein
frame and string frame becomes much more subtle.  In particular
we will claim that it is possible for an Einstein metric with
cosmological horizons to map to a string frame metric with no
cosmological horizon.

Note that we do not claim to have found a de Sitter space solution
to string theory, nor indeed any solution with cosmological horizons.
De Sitter space does not admit supersymmetry, and it
is notoriously difficult to find stable solutions to string theory
without supersymmetry
\cite{Maldacena:2000mw} \cite{Gibbons:2001wy}.  But with de
Sitter solutions, the question has been whether the cosmological
horizons form an obstruction in principle to finding a solution,
beyond the usual complexities of breaking supersymmetry.  We claim
that the appearance of a cosmological horizon is not necessarily
sufficient to doom any theory of quantum gravity, and that in
certain circumstances it is possible to avoid any difficulties by
an appropriate field redefinition.

\section{Skirting the Horizon}

The metric of a $D$-dimensional de Sitter space
(in conformal coordinates) may be written as

\be
ds^2 = {1\over cos^2 T} (-dT^2 + d\Omega_{D-1} )
\ee
where
\be
-{\pi \over 2} <T<{\pi \over 2}
\ee
The conformal factor ${1\over cos^2 T}$ is irrelevant to the
causal structure of de Sitter space.  One sees that the appearance
of cosmological horizons is a result of the limits on the
range of the conformal time $T$.  If an observer is located at
the north pole of the $(D-1)$-sphere parameterized by $d\Omega$,
then a signal from the south pole will only arrive after a
conformal time $\delta T = \pi$.  Thus a signal from the south
pole emitted at the ``beginning of time" will just reach the
north pole at the ``end of time."  A signal emitted from the south
pole any later than the $T=-{\pi \over 2}$ will never reach
the observer at the north pole.  This is the statement that a
cosmological horizon exists.
Because of the conformal factor in the metric, a particle will
reach $T= \pm {\pi \over 2}$ only after an infinite proper time.

However, the proper time necessary to reach any point
$T={\pi \over 2} -\epsilon$ before the end of time is finite.
The infinite proper time thus corresponds to a very small region
in conformal coordinate time.  Since strings are not sensitive
to distances shorter than $l_s$, one might expect
that a string at $T={\pi \over 2}-\epsilon$ would naturally
smooth out this distance and probe $T={\pi \over 2}$ as well.
The necessity for strings to probe across this boundary
would be similar to what is seen in \cite{Lowe:1995ac}.
But the resolution of this question should depend on the
metric which the string actually sees.

We will treat the de Sitter solution above as a solution in
Einstein frame.  Although the metric will be of de Sitter form,
we will consider different forms of the dilaton solution.
It is well-known that one can change the conformal
factor in the metric for a varying dilaton by going to string
frame, and that is precisely what we will do.
Using the standard transformation
between Einstein and string frame appropriate to Type IIA supergravity
\footnote{In a realistic model, de Sitter space might emerge as
a 4-D effective theory when 10-D flat space is compactified on some
non-trivial manifold.},
one finds that if the string coupling vanishes quickly enough at
the points $T={(2n+1)\pi \over 2}$, then the conformal factor
in the string frame metric will not blow up at those points.
For simplicity, we may choose the particularly simple form

\be
g = g_0 cos^4 T
\ee
in which case one would get a string frame solution of the form

\be
ds^2 = -dT^2 + d\Omega_{D-1}
\ee
\be -{\pi \over 2} < T < {\pi\over 2}.
\ee

The important thing to note is that in string frame, a particle will
reach the end of time at $T={\pi\over 2}$ in {\it finite} proper time.
There is no curvature singularity at this point; from the
$D$-dimensional point of
view everything is perfectly smooth.  As a result, there is no
obstruction to
simply extending the metric so that the conformal time $T$ runs over
all real values.  But the fact that the range of $T$ was limited was
precisely what caused the appearance of the cosmological horizons in the
first place.  In the new covering space, there are no cosmological
horizons.  The removal of the cosmological horizon by adopting a covering
space depended on the fact that one reached the end of time in a finite
proper time in string frame.

In order to obtain a de Sitter space solution in Einstein frame, one
should expect to have a stress energy tensor of the form
$8\pi T_{\mu \nu} = -\Lambda g_{\mu \nu}$, with $\Lambda$ being a positive
cosmological constant.  The contribution of the dilaton kinetic term
to the stress energy tensor for the above solution will clearly not
be homogeneous, so one would have to have additional contributions from
other scalars or field strengths which combine to yield a homogeneous
energy tensor.  For less symmetric solutions with a cosmological
horizon, one would expect this condition to be relaxed.

\section{Instaton-Like Singularity}

The vanishing of the string coupling at a locus of space-time is a
signal of the appearance of a singularity, which is often resolved
by new degrees of freedom.  From the 11-dimensional point of view,
it corresponds to a singularity which looks like a cone,
in which the size of the dilaton direction shrinks to
zero at the locus.
Another example would be certain
supergravity solutions where $e^{\phi}$ vanishes at the core.  This
corresponds to the existence of a brane at the core.  In our case,
however, the locus on which the string coupling vanishes is localized
in time, suggesting that whatever degrees of freedom appear at this
singularity are in some sense instantonic.

In order for a cosmological horizon to disappear under a change of
frame, it is necessary (but not sufficient) for the string coupling
to vanish at the ``end of time."  Thus, it seems that this potential
mechanism for avoiding the problems of cosmological horizons depends
intimately on the appearance of instanton-like singularities.

In \cite{Balasubramanian:2002ry} examples of such instanton-like singularities
were discussed.  They considered certain $Z_2$ orbifolds in which
the time coordinate as well as space coordinates were inverted under
the $Z_2$ action.  For simplicity, consider an example where only
one other spatial coordinate $x$ is inverted.  The $t$ and $x$ coordinates
together form a cone, which is singular at the tip.  By uplifiting
to M-theory and taking the angular direction on the cone to
parameterize the dilaton direction, one finds that the string coupling
constant vanishes at the instanton-like singularity, exactly as is desired
in the scenarios discussed here.  Similarly, M-branes wrapped on the
singularity yield degrees of freedom fixed to the singularity.

\subsection{Frame Dependence?}

It seems very odd indeed that something as significant as a cosmological
horizon should be frame dependent (issues relevant to the comparison of
physics in different frames were discussed in \cite{Kaloper:1997sh}).
Usually, physics in Einstein frame
and in string frame are equivalent, but it appears that this
connection may be much more subtle in the presence of instanton-like
singularities.

There is another context in which one sees a somewhat similar distinction
between Einstein and string frame.  Consider a linear dilaton solution to
bosonic (or super)string theory:

\bea
g^s _{\mu \nu} &=& \eta_{\mu \nu} \nonumber\\
e^{\phi} &=& e^{ax},
\eea
where $a=\sqrt{26-D \over 6 \alpha'}$ and the worldsheet theory
has conformal invariance.
If one changes to Einstein frame, the new metric is of the form

\be
g^e _{\mu \nu} = e^{-cx} \eta_{\mu \nu},
\ee
where $c={4a \over D-2}$.

In string frame, the proper distance between a point in the
interior and the strong
coupling region at $x=\infty$ is infinite, while in the Einstein frame
that distance is finite.  This well-understood case is actually quite
similar to the scenarios discussed here, except that we have looked at
cases where the dilaton was time-dependent instead of spatially-dependent.
As a result, we have found that the proper time between a point in the
interior and the weak coupling region is infinite in the Einstein frame,
but finite in string frame.

\subsection{New Instantaneous Degrees of Freedom}

The appearance of these new degrees of freedom behind the cosmological
horizon is reminiscent of a similar phenomenon which occurs when black
hole event horizons appear.  Indeed, the questions about how one should
formulate quantum gravity in the presence of a cosmological horizon could
just as well have been asked in the presence of a black hole event horizon.
If one forms a charged black hole which never completely decays through
Hawking radiation, then it might appear that this process cannot be described
by an S-matrix, since some of the infalling material falls behind a black
hole horizon and never emerges to reach asymptotic infinity.  But it is
believed that the solution to this problem involves a microscopic description
of the states of a black hole.  In particular, it is believed that this
microscopic description allows the states of the black hole to be included
as part of the outgoing states, and thus allows the process of matter falling
into a black hole to be described by an S-matrix.  For certain extremal
(or near-extremal) black holes, this idea can be made concrete by describing
the degrees of freedom of the black hole in terms of D-branes or strings.

It might seem that the degrees of freedom on this instanton-like object which
appears in this scenario might
similarly provide additional outgoing states, which allow string theory in
the presence of a naive cosmological horizon to still be described in terms of
an S-matrix.  But one significant difference between this case and the black
hole case is that the strings and branes which described the microscopic black
hole states were extended in time.  Thus, one could easily see how they could
contribute to the set of outgoing states at future infinity.  But these new
instanton-like degrees of freedom are fixed at a point in time at the edge
of each de Sitter patch.  It is not obvious that they can be ``seen" in the
asymptotic future.  It was suggested in \cite{Balasubramanian:2002ry}
that these states may have to be traced over when computing scattering
amplitudes.

\section{Conclusions}

It is not clear what to make of this apparent distinction between the
physics of the Einstein frame and the string frame in the presence
of these instanton-like singularities.  This could possibly be an indication
that particle physics is a bad approximation in the presence of these
instanton-like singularities and cosmological horizons.  Instead, space-time
is inherently ``stringy."  The idea that the particle approximation gives
way to a string formulation is not new; it is known that Einstein
gravity is unrenormalizable as a field theory, and must give way to some
more fundamental theory in the ultraviolet (with string theory being the
most likely candidate at the moment).  But we seem to have have evidence
that the particle approximation can break down in the infra-red as well.
One can probe intermediate distance scales in these backgrounds with
particles by performing scattering experiments in a background which is
similar at the desired scale, but which has a vanishing energy density at
larger scales (so that there is no cosmological horizon).  But if the
region to be probed is large enough, then a cosmological horizon will
appear in the interaction region
and the particle approximation will break down.
It may be necessary instead to treat the objects to be scattered as
objects in string theory (or M-theory).  In the presence of certain
instanton-like singularities, it seems that this description does provide
for a meaningful S-matrix.  Since the string frame metric is smooth
with no horizons throughout the covering space, there seems to be no
obstruction to strings which stretch from one de Sitter sheet to
another (although this cannot be made explicit without a precise
solution).  Apparently string fields which are located outside
each others putative cosmological horizons can nevertheless be
correlated, in a sense similar to \cite{Lowe:1995ac}.

Unfortunately, these arguments are extremely heuristic and somewhat
speculative.  Clearly, this is not presented as a solution to the
problems of cosmological horizons.  Rather, it is a potential
scenario which could be useful in formulating such a solution.
It would be
interesting to generate backgrounds with such instanton-like singularities
and cosmological horizons, and to determine if all solutions to M-theory
with cosmological horizons also have instanton-like singularities which are
covered by the horizon.

\vskip .2in
{\bf Acknowledgments}

We are grateful to J. Brodie, B. Grinstein, E. Halyo, S. Hellerman,
K. Intriligator, S. Kachru, N. Kaloper, A. Manohar,
J. McGreevy and E. Silverstein for useful discussions.  This
work is supported by
DOE-FG03-97ER40546.

\end{document}